\documentclass[a4paper,fleqn]{cas-dc}
\usepackage[switch]{lineno}
\usepackage[numbers]{natbib}
\usepackage{nccmath}
\usepackage{lineno}
\usepackage{float}
\usepackage{amsmath}
\usepackage{parskip}
\usepackage{caption} 
\usepackage{upgreek} 
\usepackage[section]{placeins}

\begin{document}
\let\WriteBookmarks\relax
\def\floatpagepagefraction{1}
\def\textpagefraction{.001}
\shorttitle{Precise Photographic Monitoring of MEG II Thin-film Muon Stopping Target Position and Shape}
\shortauthors{D. Palo, W. Molzon et~al.}

\title [mode = title]{Precise Photographic Monitoring of MEG II Thin-film Muon Stopping Target Position and Shape}                

\author[1]{D. Palo}[orcid=0000-0001-9256-5348]

\author[2]{M. Hildebrandt}
\author[2]{A. Hofer}
\author[1]{W. Kyle}
\author[1]{D. Lad}
\author[1]{T. Libeiro}
\author[1]{W. Molzon}[orcid=0000-0001-9232-6064]
\cormark[1]
\ead{wmolzon@uci.edu}

\address[1]{University of California, Irvine, CA 92697}
\address[2]{Paul Scherrer Institut PSI, 5232 Villigen, Switzerland}

\cortext[cor1]{Corresponding author}

\begin{abstract}
We describe and show results of a photographic technique for continuously monitoring the position, orientation, and shape of a thin-film muon stopping target for the MEG II experiment\cite{MEG II}. The measurement is complicated by the target being located in a region with 1.3 T magnetic field, significant radiation and having limited access. The technique achieves a measurement precision of $10$ $\mathrm{\upmu m}$ normal to and $30$ $\mathrm{\upmu m}$ parallel to the film surface, significantly better than required for the MEG II experiment.  
\end{abstract}

\begin{keywords}
\vspace{-6px}
Photographic Position Monitoring \vspace{-6px} \sep Image Analysis \vspace{-6px} \sep Charged Lepton Flavor Violation \vspace{-6px} \sep Muon Stopping Target

\end{keywords}

\maketitle

\section{Introduction}
The MEG II experiment\cite{MEG II} is a search for the decay of an anti-muon ($\upmu^{+}$) to a positron ($e^{+}$) and a photon ($\gamma$): $\upmu^{+} \rightarrow e^{+} +\gamma$.  This is an example of a process involving charged leptons that violates additive quantum numbers associated with muon and electron number (CLFV). No example of CLFV has been seen, and an observation of any CLFV process would have profound implications for our understanding of the fundamental constituents of matter and how they interact.

The experiment proceeds by stopping $\upmu^{+}$ in a thin plastic film (the stopping target) inside a superconducting solenoid with magnetic field of  1.3 T. The $e^{+}$ and $\gamma$ are detected in a magnetic spectrometer and a fully absorbing calorimeter, respectively. Backgrounds that might fake the signal are rejected by precisely measuring the momenta and times of the $e^{+}$ and $\gamma$; true signal events have $e^{+}$ and $\gamma$ originating from the stopping target at the same time, with equal magnitude of momentum and direction back-to-back. 
The primary sources of background do not have these characteristics.

We describe here a technique to increase the precision of the $e^{+}$ kinematic measurements at the decay vertex on the target by precisely measuring the position, orientation, and shape of the stopping target with respect to the magnetic spectrometer. The $e^{+}$ direction is determined by projecting the helical trajectory measured in the spectrometer to the target plane. An error in the position of the target in the direction normal to the target plane would result in an error in the $e^{+}$ direction due to the incorrect path length to the target and hence incorrect amount of curvature. The precision with which the target position must be measured is set by the requirement that the impact of any error on the $e^{+}$ direction be less than that of other contributions to the error in the relative $e^{+}\gamma$ angle. For MEG II, this angle will be measured with a precision of $< 6$ mrad and the goal is for the uncertainty in the $e^{+}\gamma$ angle due to uncertainty in the target plane position to be $< 0.6$ mrad. This corresponds to an error in the target position or shape of 85 $\mathrm{\upmu m}$ normal to the target surface, giving a path length error of 120 $\mathrm{\upmu m}$ for a 53 MeV/c $e^{+}$ incident at $45^{\circ}$ with respect to the film's surface in a 1.3 T magnetic field.  

Experience with a similar target in the MEG experiment\cite{MEG} showed that the target shape changed over a period of $\sim 1$ year of operation, developing a bowing with maximum deviation from the plane of approximately 1 mm. The time dependence of the bowing was not well monitored. Possible time dependent target motion might also result from the periodic pneumatically actuated extraction and insertion of both the MEG and MEG II targets; this is done for the purpose of acquiring special data used to calibrate the detectors. These two time-dependent effects motivate the requirement to monitor the position, shape and orientation of the target continuously. 

 The previous most sensitive search for $\upmu^{+} \rightarrow e^{+} +\gamma$, done by the MEGA experiment\cite{MEGA}, also required a precise measurement of the position and orientation of a stopping target (in this case 0.1 mm thick Mylar). The target experienced deformations of $\pm 1$ mm normal to the target. "The position of the target when mounted in the spectrometer was determined by direct visual measurements, based on a grid penned on the target surface"\cite{MEGA}. The experiment fit the measured positions (approximately 100) to a plane defining the target's position with "errors of 1 mm on the spacial points"\cite{MEGA}.

The photographic technique described in this paper monitors continuously the \textit{change} in the target's shape and in its position and orientation with respect to the camera. It does this by imaging approximately 120 dots printed on the stopping target using a camera located about 1.2 m from the target. This technique will not measure the absolute position of the target. To determine the position and orientation of the target with respect to the magnetic spectrometer, MEG II will use two independent techniques to correlate the target position in the camera coordinate system to the position with respect to the magnetic spectrometer. Additionally, bench measurements will characterize the target shape.
The first technique is an optical survey, done only infrequently, of the target with respect to the magnetic spectrometer. The correlation with the photographic results is done by analyzing a sequence of very precise photographic measurements taken simultaneous with the optical survey. The second technique (described in Appendix \ref{app_tracking}) was first used in the MEG experiment\cite{MEG}; it uses momentum analyzed positrons recorded during data-taking to image, in 3 dimensions, small holes in the target by detecting a deficit of positrons originating from the position of the holes. The MEG experiment achieved a precision in the measurement of the target position normal to its surface of $0.3 -0.5$ mm. The uncertainty was primarily due to lack of statistics available to measure the time dependence of the position. By correcting for time dependence in the target geometry using the photographic technique, the full data-set of MEG II will be used for measuring relative alignment of the spectrometer and the target. This second technique has an additional advantage that it is not affected by possible errors in the optical survey. 

In the remainder of this paper we describe only the camera system, its operation, and analysis of images, including results of operations at full beam. Neither an optical survey of the target nor acquisition of momentum analyzed positron data has yet been done.   

\begin{figure*}[!b]
{\centering
\includegraphics[width=16cm]{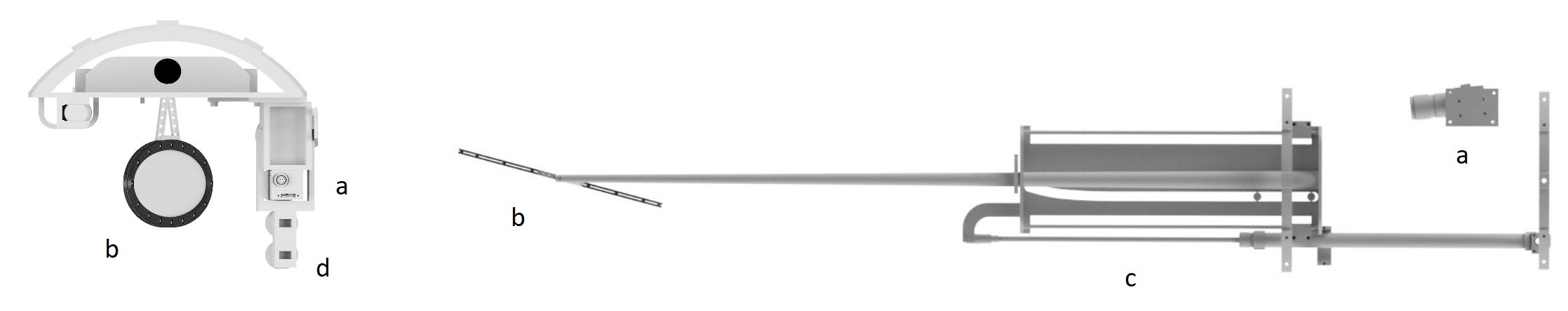}
\caption{Drawings of the assembly including a) the camera used in this technique, b) the stopping target, c) the pneumatically actuated target support, and d) the LED lights. The camera is positioned at the vertical (MEG II y-axis) center of the target, and is rotated to center the target in the image plane. Left: Beam-axis view of assembly. Right: Top view of the assembly, with some support pieces, a second camera not used in this analysis, and cabling omitted for clarity. 
\label{draw}
}}\end{figure*}

\section{Methods}
\subsection{Camera Installation and Operation}

The implementation of the photographic alignment system is complicated by several factors. Access is limited since the stopping target is at the center of a tracking detector $\sim $ 2 m long in a solenoid with nominal field of 1.3 T at the target location. No simple optical path from the target to a position outside the solenoid exists. The camera system cannot be closer than $\sim $ 1.2 m from the target, with the camera axis nearly along the magnet axis (Figure \ref{draw}), and it must be located at the incoming muon beam end of the spectrometer. The magnetic field at this location is $\sim $ 0.8 T. Further, there is significant radiation at the camera location, primarily from positrons from muon decay in the target. This presents the possibility of radiation damage to the sensor and camera electronics. Finally, the available space for the camera and lens is limited. 

The camera is mounted approximately 1.2 m from the target, at the same height as the target, and offset horizontally from the target by approximately 10 cm (Figure \ref{draw}). The camera axis is at an angle of $5.60^{\circ}$  with respect to the MEG II magnet axis and $69.40^{\circ}$ with respect to the vector normal to the target plane. The camera is mounted to a support structure that is rigidly attached to a spool piece attached to the cryostat of the COBRA (COnstant Bending RAdius)\cite{COBRA} magnet. The structure also supports two LED lamps for illuminating the target, another camera not used for the position monitoring described here, and a pneumatically controlled target support that moves the target between the inserted position when it is being used and an extracted position during certain calibration data taking. 

The positron spectrometer is also attached to the COBRA magnet cryostat. To monitor any possible motion of the spectrometer with respect to the camera, we will image features printed on the spectrometer structure. For the tests described here, we verified that there was no motion of the target with respect to the spectrometer by imaging and monitoring the position of a flange on the spectrometer structure.

The alignment system is implemented with an industrial \linebreak camera\cite{Imaging}  with a 1/2.3-inch CMOS sensor with $3856 \times 2764$ pixels, each 1.67 $\mathrm{\upmu m}$ square, and a 50 mm lens. The interface to an acquisition computer is by USB3; Ethernet interfaces do not work in the 0.8 T magnetic field. We use an active USB extender cable\cite{Extension} to allow the acquisition computer to be at sufficiently small magnetic field. The manufacturer provides a graphical user interface and provisions for operating the camera with scripts that can be written in C++ or Python\cite{pyicic}. The software provides the capability to set the frame rate, exposure, gain, and fraction of the image plane to be read out. The focus and aperture are set manually. 

The camera control and image acquisition are implemented using Python scripts combined with scripts to control the LED lights and sequencing of image acquisition. All the acquisition and control software is integrated with the MEG II data acquisition and control system. During operation, we acquire a dark field image and a set of three normal images every 15 minutes. The dark-field image is used to subtract the background intensity in \lq{hot}\rq pixels. The hot pixels remain in fixed position and comprise 0.05\% of all pixels. They are consistent with being caused by radiation damage as they appeared only following significant camera use during operation of the beam.

\subsection{Target Construction}

\begin{figure}[H]
{\centering
\includegraphics[width=8cm]{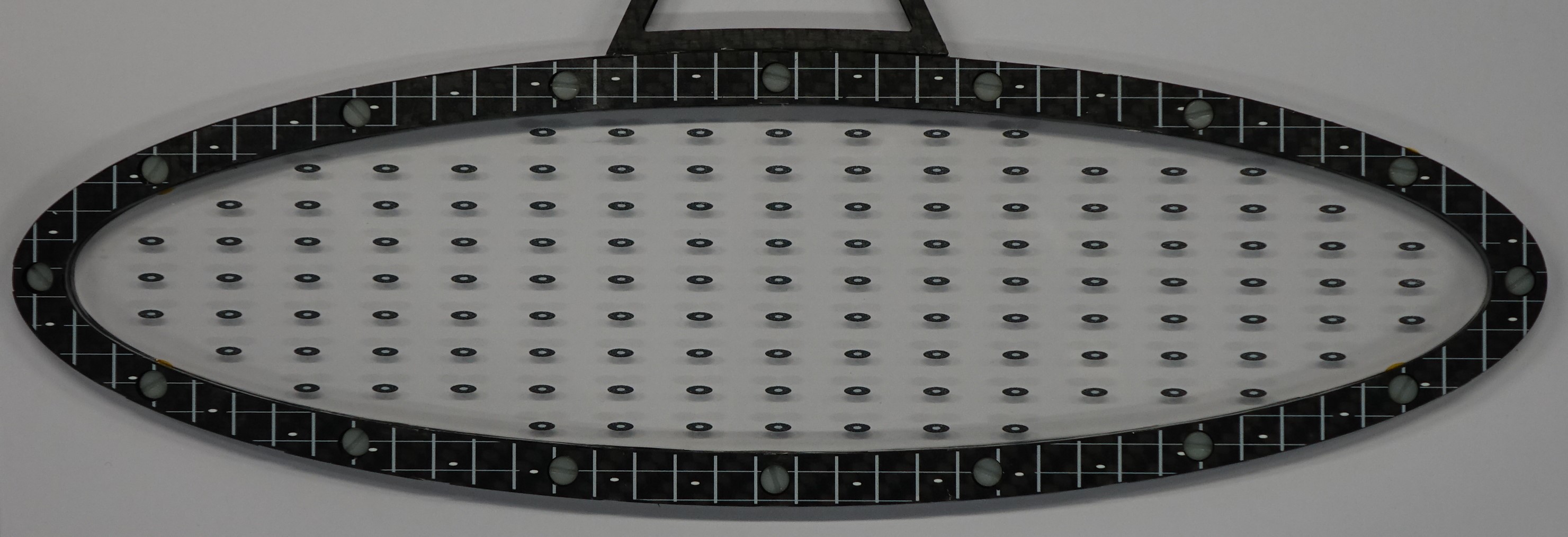}
\caption{Head-on image of the target.}
\label{target}
}

\end{figure}

The MEG II stopping target is made of a thin scintillating plastic film with average thickness of $174$ $\mathrm{\upmu m}$
ranging from 155-194 $\mathrm{\upmu m}$ and elliptical in shape, with width 270 mm and height 66 mm (shown in Figure \ref{target}). It is supported between two hollow carbon-fiber box frames and allowed to float to avoid stress on the target due to dimensional changes in the frames or foil. An array of white dots, each superimposed on a black background, is printed on both the frame and the film. The dots are elliptical with width 1.52 mm and height 0.51 mm on the film, and 1.27 mm and 0.42 mm on the frame such that the dots appear circular when imaged at an oblique angle with respect to the target's surface.

\subsection{Image Analysis and Dot Characterization} \label{image_analysis}

\begin{figure}[H]
{\centering
\includegraphics[width=8cm]{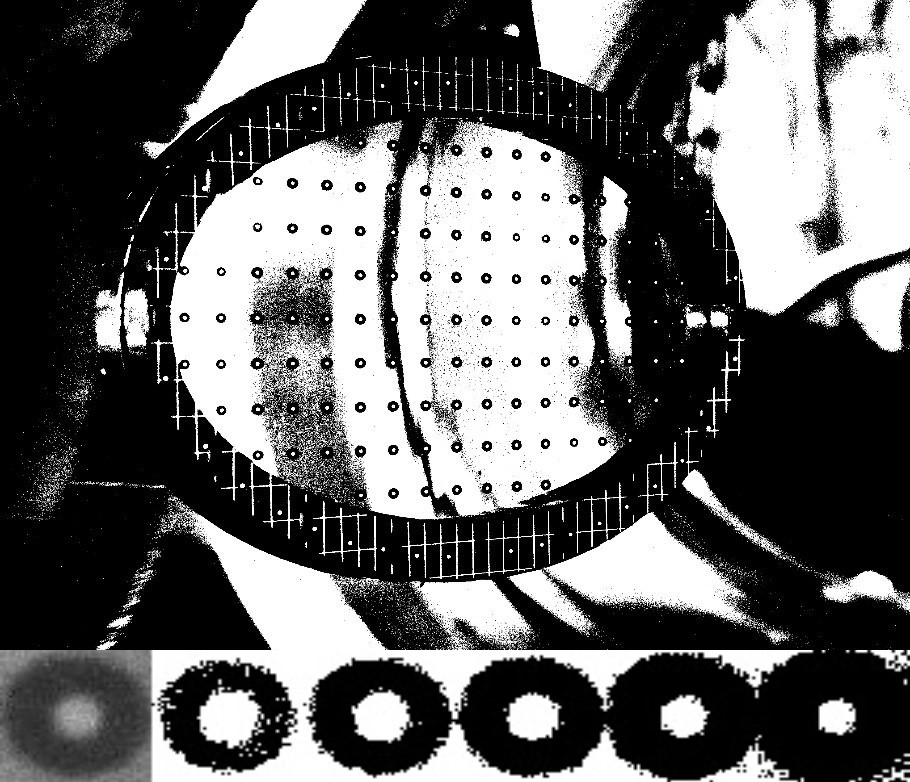}
\caption{Top: A typical binary image of the target\cite{OpenCV_binary}. Bottom: From left to right we show an original (non-binary) dot followed by binary versions with increasing threshold parameter values.  
\label{binary}
}}\end{figure}

In this section, we describe the image analysis to determine the position of each dot in the image plane using an open-source code\cite{OpenCV}. First, the code\cite{OpenCV_binary} produces a binary version of the image, such that all pixels with an intensity above (below) a selected threshold turn white (black). The binary version is shown in Figure \ref{binary} for a threshold pixel intensity of 80 (out of full scale value of 255). The image shows that white dots on both the frame and the film are distinguished from the black background. 

We then apply code\cite{OpenCV_contours} to the binary image to find clusters of white pixels. The code calculates parameters for each cluster such as the cluster's moments, size, and aspect ratio. We associate the clusters with target dots, eliminating false positive clusters based on selection criteria such as the cluster's size, aspect ratio, and expected dot positions based on the printed pattern. Specifically, by using the printed pattern, we create a 2D grid on the image such that one target dot is expected in each grid location. If only one dot is found in the bin, the dot is assigned a unique 2D index, otherwise no dot is assigned to that index. The procedure finds dots with efficiency of $>$95 \% even with large variation in background lighting. To avoid the cluster's centroid depending on details of pixels association near the cluster's edge, each dot's centroid is calculated using an intensity weighted mean pixel position. This method determines the centroid of each dot with a dispersion of $\sigma = 0.2$ $\mathrm{\upmu m}$ at the image plane (approximately 4.8 $\mathrm{\upmu m}$ at the target) determined from a series of sequential images taken close in time. 

We have verified that the lighting intensity and the threshold parameter do not affect the measured positions in a systematic way, although they do affect the cluster size (Figure \ref{binary}). The dispersion in the measured dot positions does marginally increase with decreased lighting. Additionally, we analyzed approximately 20 images while varying the threshold parameter; this did not have a systematic effect on the resulting fit for the target shape, position, and orientation (see Appendix \ref{app_thresh}).

\subsection{Calculation of Target Geometry} \label{geo_sect}

 The dot coordinates determined in Section \ref{image_analysis} are next used to fit for changes in the target's position, orientation, and shape.

The analysis described here is done in a coordinate system aligned with the camera. It is defined such that the origin is at the camera lens' position, the z-axis aligns with the camera's optical axis, the y-axis is vertically upward, and the x axis defines a right-handed coordinate system. This coordinate system is nearly aligned with the MEG II system, which differs only in the z axis being aligned with the incoming muon beam. The precise transformation from the camera coordinate system to the MEG II coordinate system will be done using a set of images taken effectively simultaneously with the optical survey of the target.

 The analysis procedure is as follows. Nominal positions and orientations of the camera and target are initially used along with the nominal positions of the dots on the target to calculate the coordinates of the $i^{th}$ dot in the camera coordinate system, given by the 3-vector $\vec{x}^{C}_{i}$. These coordinates are projected to the camera image plane using camera optics, yielding the {\it nominal} image plane coordinates ($u^{C}_{i}$, $v^{C}_{i}$). These projected image plane coordinates are then compared to the measured image plane coordinates ($u^{I}_{i}$,$v^{I}_{i}$). The best fit target location in the camera coordinate system is then determined by minimizing the sum of the squares of the residuals of the measured and projected image plane dot coordinates, varying the parameters describing the target position, orientation, and shape.

The projection of 3D coordinates is {\it nominally} done using the focal length approximation (Equation \ref{pre_focal}), and the magnification equation (Equation \ref{pre_mag}). In these equations, \textit{efl} is the effective focal length, $d_{o}$ is the object distance, $d_{i}$ is the image distance, $h_{o}$ is the object height, and $h_{i}$ is the image height. 

\begin{equation}
\frac{1}{e\!f\!l}= \frac{1}{d_{i}} + \frac{1}{d_{o}}
\label{pre_focal}
\end{equation}

\begin{equation}
M=\frac{h_i}{h_o}=-\frac{d_i}{d_o}
\label{pre_mag}
\end{equation}

Here we discuss two shortcomings of the nominal projection that we correct to define better the camera optics.  First, the effective focal length \textit{efl} is known only approximately from manufacturer information. Therefore, for a fixed $d_{o}$ and $d_{i}$, the magnification is only approximately known. Second, since the object distances vary by approximately 20 cm, the edges of the target are not in focus, and thus the focal length approximation only approximately calculates the value of $d_{i}$ and therefore the magnification for the out of focus part of the target. To address the second issue, we perform a ray tracing program for varying object distances to determine the position of each dot on the camera's image plane.  For in-focus object distances of 120 cm and a fixed lens position with an effective focal length of 50 mm, an object 10 cm out-of-focus has a magnification error of $\sim 0.4$ \% from the focal length approximation. We have verified that the ray tracing program is equivalent to a corrected magnification equation, given below. 

\begin{equation}
M=\ \frac{h_i}{h_o}=\ -\frac{I}{d_o} 
\label{mag_eq}
\end{equation}

Here $I$ is defined as the {\it fixed} distance from the center of the lens to the camera's image plane ($d_{i}$ for the in focus object). To address the first shortcoming (determining the camera's effective focal length), we optimize the distance $I$ using a procedure described in Appendix \ref{app_optical}). With these corrections applied, the projection is given below. 

\begin{equation}
u^{C}_{i}=-I\frac{x^{C}_{i}}{z^{C}_{i}} \ , \  v^{C}_{i}=-I\frac{y^{C}_{i}}{z^{C}_{i}}
\label{projection_eq}
\end{equation}
Using the corrected projection equation, we minimize the $\chi^{2}$ defined below, varying the target position and orientation (6 parameters) and one or more parameters for target plane distortions. Here, $\sigma =0.12$ pixels based on the dispersion (discussed in Section \ref{image_analysis}) of the dot coordinate measurements.  

\begin{equation}
\chi_{1}^{2}= \mathrm{\Sigma}_{i}\ \frac{(u^{C}_{i}-u^{I}_{i})^{2} + (v^{C}_{i}-v^{I}_{i})^{2}}{\sigma^{2}} \label{chi1_eq}
\end{equation}

The seven parameters that minimize the $\chi^{2}$ rotate, translate and deform $\vec{x_{i}^{C}}$ (the 3D dot coordinates in the camera reference frame) \textit{prior} to the projection. The parameters are three translations along the camera coordinate system axes ($\vec{x^{C}}$) and three rotations about these axes. Since the rotation angles are very small, "the sequence of rotations is unimportant"\cite{Goldstein}. The rotation is described by the matrix below, where $c_{i}$ and $s_{i}$ abbreviate cos(i) and sin(i), and $\phi$, $\theta$, and $\psi$ represent rotations about camera's axes $x^{C}$, $y^{C}$, $z^{C}$ respectively. 

\begin{equation}
 \begin{pmatrix} 
c_{\theta} c_{\psi} & s_{\phi}  s_{\theta}c_{\psi} - c_{\phi} s_{\psi}& c_{\phi}s_{\theta}c_{\psi} + s_{\phi} s_{\psi}\\
c_{\theta} s_{\psi} & s_{\phi}s_{\theta}s_{\psi}+ c_{\phi}c_{\psi}&c_{\phi}  s_{\theta}s_{\psi} - s_{\phi} c_{\psi}\\
-s_{\theta}& s_{\phi} c_{\theta} & c_{\phi}c_{\theta} \label{rotation}
\end{pmatrix}
\end{equation}

We fit for a global deformation normal to the surface, treating the film's surface as a paraboloid, restricted such that the film's deformation is null at its perimeter. The deformation parameter is defined as the maximal deformation normal to the target (located at the target's center). The equation for the bowing parameter, c, is given below.

 \begin{equation}
 x^{T}_{i} =  c(\frac{z^{T}_{i}  - z_{0}^{T}}{a})^{2} + c(\frac{y^{T}_{i}  - y_{0}^{T}}{b})^{2} - c \label{bow_eq}
 \end{equation}
 
In Equation \ref{bow_eq}, the coordinates are in the target coordinate system, where the origin is at the target center, the vector normal to the target plane defines the x-axis (most aligned with the x-axis of the MEG II coordinate system), the y-axis is vertically upward, and the z-axis is along the long axis of the target. The parameters $z_{0}^{T}$ and $y_{0}^{T}$ define the center dot's position, and a and b are the target's semi-major and semi-minor axes respectively.

The $\chi^{2}$ in Equation \ref{chi1_eq} is minimized using the Nelder-Mead method (using open source code\cite{Scipy}). We compared the method with the Powell method\cite{Scipy}; they find the same minimum. The $\chi^{2}$ minimization yields the optimal seven parameter transformation, which is then applied to the 3D dot coordinates, $\vec{x^{C}}$, thus calculating the target's position, orientation, and shape with respect to the camera.

In addition to the parameters describing the target's geometry, the minimization code produces a two-dimensional image plane residual for each dot (e.g. $u^{Res}_{i} = u^{C}_{i} - u^{I}_{i}$).
Ideally, these residuals should be random and consistent in size with the precision with which the dot coordinates are found. However, they also contain information such as effects from higher order target deformations, printing offsets, and film irregularities that are not captured in the dot coordinate measurement precision. These latter effects will be captured in bench measurements of the target; these have not yet been done. Specifically, the transverse position of each dot position on the target will be measured; any residual will be treated as a deformation normal to the target surface as we assume that any target deformation predominantly moves the surface perpendicular to the target plane (no stretching in the plane). In this way we can characterize {\it any} deformation or foil irregularity normal to the surface at the position of each dot, no longer restricting the 3D dot positions to parabolic deformations. The precise way in which the foil deformation will be analyzed will be better informed when the target is better characterized.

 \begin{figure*}[!ht]
{\centering
\includegraphics[width=16cm]{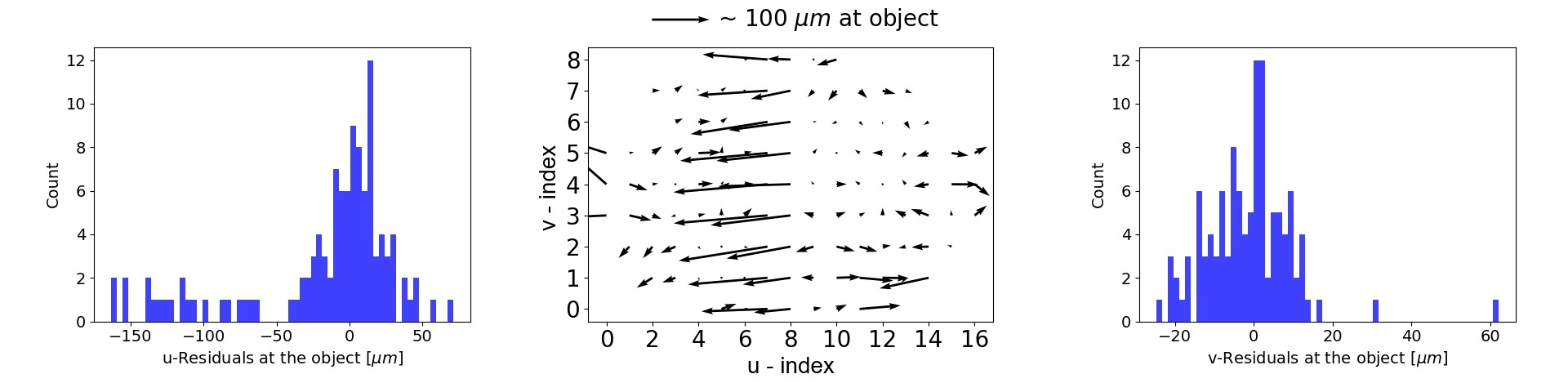}
\caption{Left: This is a histogram of  the u residuals. The dots entries with large negative residuals were removed from the fit to get the baseline target geometry. Middle: This is a scatter plot of the vector of the average u-v residuals for each dot. Right: This is a histogram of the v residuals. \label{avg_res}
}}\end{figure*}

In order to establish a baseline nominal target geometry without having characterized the target, we fit 20 sequential images taken close in time and calculate the average residuals for each dot. Dots with large (>$10 \sigma$) residuals (presumably due to offsets of the type described above) are removed and the fit is repeated to determine the seven parameter coordinate transformation. This is applied to \textit{all} dots to determine their coordinates in the camera coordinate system. In the final implementation, this procedure will be done using images taken simultaneous with the optical survey in order to establish the correspondence between location of the target in the camera and MEG II coordinate systems. These dot coordinates are then projected to the image plane, and the residuals are calculated (shown in Figure \ref{avg_res}).  The v-residuals($v^{Res}$) are randomly distributed with a dispersion of $\sigma = 0.45$ $\upmu m$ at the image plane. Over much of the target, the u-residuals($u^{Res}$) are also small and randomly distributed, but have large negative values for column indices seven and eight ($\langle u^{Res} \rangle$ = -5.33 $\upmu m$  at the image plane). We have confirmed that these large systematic residuals are not due to printing offsets by measuring the distance between the columns of dots at the image plane from a head-on image (Figure \ref{target}). They are seen as incorrect spacing of columns in the images taken at the oblique angle used to calculate the residuals. The residuals are consistent with a deformation of the target normal to the target surface of $ \sim 100$ $\upmu m$. This deformation has not been confirmed with bench measurements of the target's shape.

Since the objective is to look for {\it changes} in the target's position, orientation, and shape, we include a correction for these average residuals at the image plane in the $\chi^{2}$. The correction is included in the modified chi-squared defined in (Equation \ref{chi2_eq}).
 
\begin{equation} 
\chi_{2}^{2}= \mathrm{\Sigma}_{i}\ \frac{(u^{C}_{i}-u^{I}_{i} - \ \langle u^{Res}_{i} \rangle)^{2} + (v^{C}_{i}-v^{I}_{i} - \langle v^{Res}_{i} \rangle)^{2}}{\sigma^{2}}
\label{chi2_eq}
\end{equation}

\begin{figure}[H]
{\centering
\includegraphics[width=8cm]{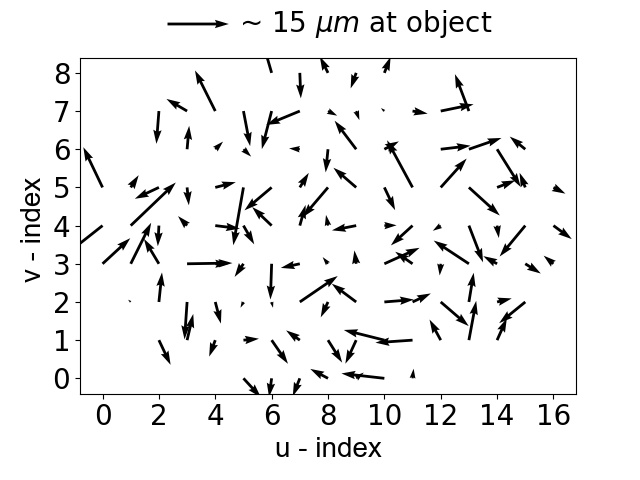}
\caption{This is a scatter plot of the vector residual for each dot using an image taken $1$ week after the fits used to characterize the nominal target position, orientation, and deformation.
\label{rand_res_eq}
}}\end{figure}

In the implementation of this technique, only images with at least 100 dots satisfying selection criteria given in (Section \ref{image_analysis}) are used. Additionally, dots with a residual larger than 4 $\sigma$ ($\sigma = 0.12$ pixels) following a first fit are eliminated and the fit is repeated.  An example residual plot with the residual correction included is shown in Figure \ref{rand_res_eq}; the residuals are now small and randomly distributed. The typical value is now $\sim 6 $ $\mathrm{\upmu m}$ at the object ($\sim 0.14$ pixels at the image plane), consistent with the dispersion of individual dot position measurements (Section \ref{image_analysis}). 

\section{Results}
\subsection{Measurements of Target Motion}

We next discuss results for measurements of target position, orientation, and deformation changes using target images acquired every 15 minutes for multiple periods of 10s of hours. 

\begin{figure*}[!hb]
{\centering
\includegraphics[width=16cm]{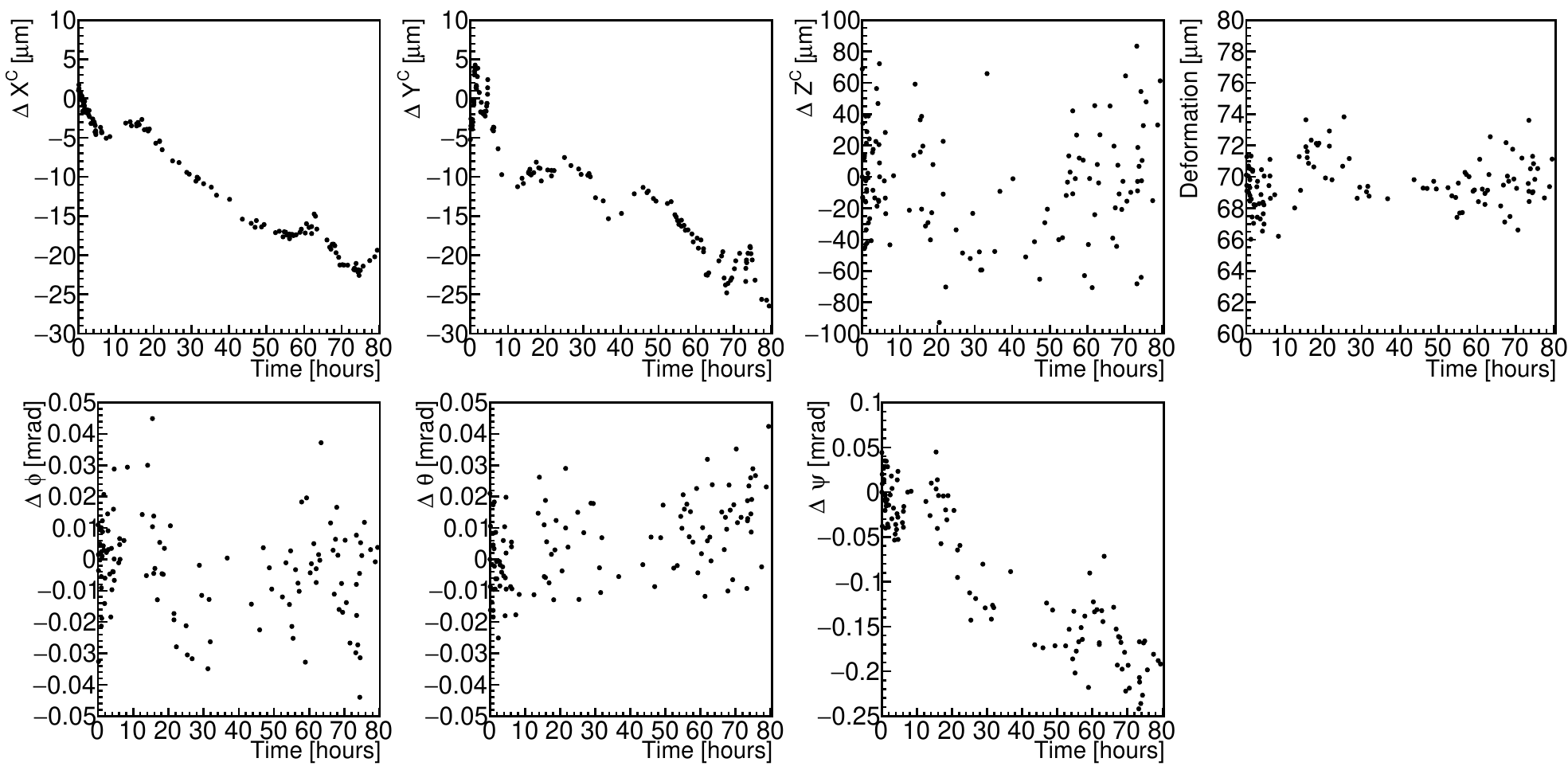}
\caption{Changes in the rigid body parameters defining the target film are plotted in the first six graphs; the seventh plots the deformation parameter.  The rotation angles are the angles defined in Section \ref{geo_sect} and correspond to rotations about the axis of the graph above (in the camera coordinate system). The largest translation is $\sim 30$ $\mathrm{\upmu m}$ in $y^{C}$ and the largest rotation is $\sim 0.3$ mrad about the camera axis, approximately aligned with the insertion-extraction mechanism. 
\label{long}
}}\end{figure*}

\begin{figure*}[!hb]
{\centering
\includegraphics[width=16cm]{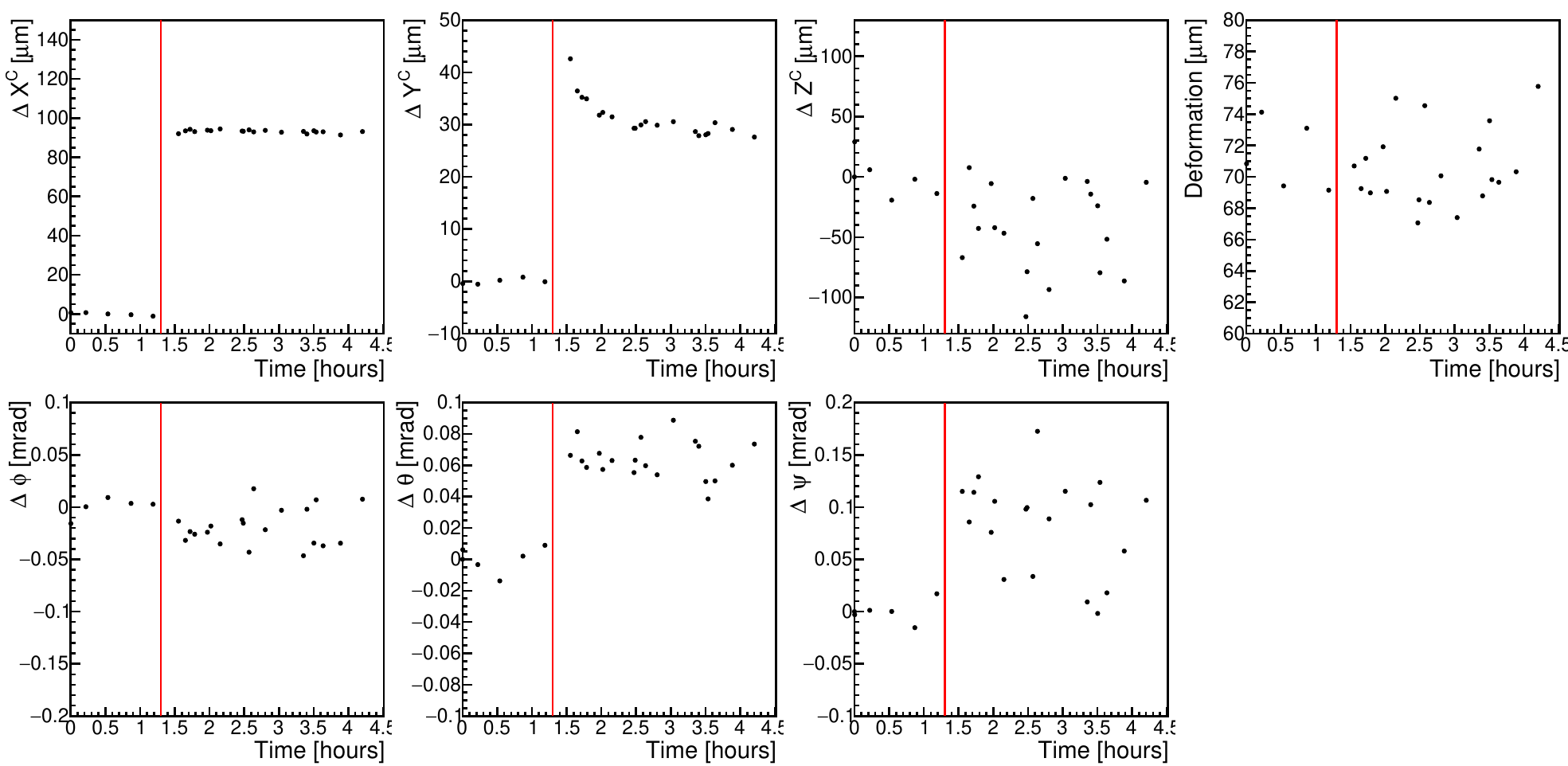}
\caption{Changes in the rigid body parameters defining the target film are plotted around the time of the insertion-extraction sequence. The rotation angles are the angles defined in Section \ref{geo_sect} and correspond to rotations about the axis of the graph above. The largest translation is $\sim 100$ $\mathrm{\upmu m}$ in $x^{C}$. The vertical line indicates the time of the extraction-reinsertion sequence.
\label{discrete}
}}\end{figure*}

During this time, two types of target motion were seen. The first was a slow drift in the target's position and orientation, on the order of 10 $\mathrm{\upmu m}$ per day (Figure \ref{long}). This is consistent with motion of the target insertion mechanism and support after the target has been reinserted. There are other $ 10$-hour periods during which the target's geometry remains stable to the precision of this analysis technique.  The second type of target motion was a discrete change in position at an extraction-reinsertion sequence; the motion is shown in Figure \ref{discrete}. This motion is a translations on the order of 100 $\upmu m$.

\FloatBarrier
\clearpage
\subsection{Precision of The Measurement of Translations Normal to the  Target Surface}
The MEG II experiment is particularly sensitive to translations of the stopping target in the direction normal to its surface ($x^{T}$).  Due to this sensitivity, we discuss three ways of calculating the variance in $x^{T}$. 

Translations in $x^{T}$ are a linear combination of translations in $x^{C}$ and $z^{C}$:  $d x^{T} = \alpha d x^{C} - \beta d z^{C}$ with $\alpha \sim  0.94$ and $\beta \sim 0.34$. The uncertainty in $x^{T}$ is, however, dominated by the uncertainty in $z^{C}$. This is because a translation in $x^{C}$ produces a uniform translation of each dot on the image plane; hence the uncertainty is proportional to the mean uncertainty in the dot position averaged over $\sim $100 dots; $\sigma_{x^{C}} \sim 0.5$ $\upmu m$, and hence a negligible contribution to the uncertainty in $x^{T}$. The measurement of $z^{C}$, on the other hand, is given to first order by the fractional change in magnification (fractional change in distance between dots at the image plane) times the distance from the camera to the target; hence the uncertainty is larger than that of $x^{C}$ by a factor equal to the distance from the camera to the target divided by the distance between dot pairs. The $z^{C}$ uncertainty is approximated by the independent measurement shown in Section \ref{indepz_section}; here the uncertainty is $\sigma_{z^{C}} \sim 50$ $\upmu m$.

The first calculation of the variance is from the dispersion in the parameters for sequential images with minimal target motion; the dispersion is shown in the first row of Table \ref{stddev_table}.  The second calculation uses an additional python package\cite{LMFIT} to calculate the covariance and correlation matrices (the correlation matrix shown in Table \ref{cor}) for the fit. The square root of the covariance matrix along the diagonal corresponds to the standard deviation in the parameters, these are displayed in the second row of Table \ref{stddev_table}. Using the value of $Cov_{[x^{C}, z^{C}]}$ (proportional to $Cor_{[x^{C}, z^{C}]}$), we calculate the uncertainty in $x^{T}$ to be $\sigma_{x^{T}} =$ 11.5 $\mathrm{\upmu m}$ and 9.6 $\mathrm{\upmu m}$ when calculated using the dispersion and the covariance matrix respectively. 

\FloatBarrier
\begin{table}[h]
\caption{Top: Uncertainty in fit parameters calculated as the dispersion in the values from a sequence of fits to images taken close in time. Bottom: Uncertainty in fit parameters calculated from the covariance matrix averaging over 100 images. The two measurements show good agreement.} 
\centering 
\scalebox{0.75}{
\begin{tabular}{c c c c c c c}  
\hline\hline
$\phi$[mrad] & $\theta$[mrad] & $\psi$[mrad]  & $x^{C}$[$\mathrm{\upmu m}$]  & $y^{C}$[$\mathrm{\upmu m}$]  & $z^{C}$[$\mathrm{\upmu m}$]  & Bow[$\mathrm{\upmu m}$]  \\ [0.5ex] 

\hline 
0.02&  0.01 & 0.03 & 0.71 & 0.61 & 39.13 & 1.00 \\ [1ex]  
\hline 
 0.01 & 0.01 & 0.02 & 0.57 & 0.40 & 29.19 & 1.13 \\ [1ex]
 \hline
\end{tabular}}

\label{stddev_table}
\end{table}

\begin{table}[h]
\caption{The table gives the values of the elements of the correlation matrix for a representative fit. } 
\scalebox{0.6}{
\begin{tabular}{c c c c c c c c} 
\hline\hline 
 & $\phi$[mrad]& $\theta$[mrad] & $\psi$[mrad]  & $x^{C}$[$\mathrm{\upmu m}$]  & $y^{C}$[$\mathrm{\upmu m}$]  &$z^{C}$[$\mathrm{\upmu m}$]  & Bow[$\mathrm{\upmu m}$] \\ [0.5ex] 
\hline 
 $\phi$[mrad] & 1 & 0.015  & -0.024 & -0.193 & -0.115 & 0.046  & -0.186 \\ [1ex] 

$\theta$[mrad]& 0.015 & 1 & -0.024 & 0.073 & 0.114 & 0.858  & 0.032 \\ [1ex] 
 
$\psi$[mrad] & -0.024&  0.073 & 1 & -0.227 & -0.059 & 0.01 & -0.219\\ [1ex] 

 $x^{C}$[$\mathrm{\upmu m}$] & -0.115  & 0.073 & -0.227 & 1 & 0.01 & -0.050 & 0.732  \\ [1ex] 
 $y^{C}$[$\mathrm{\upmu m}$] & -0.115 & 0.114 & -0.059  & 0.01 & 1 & 0.131  & 0.01 \\ [1ex] 
 
$z^{C}$[$\mathrm{\upmu m}$] & 0.046 & 0.858 & 0.01 & -0.050 &  0.131  & 1  & -0.051\\ [1ex] 

Bow[$\mathrm{\upmu m}$]  & -0.186 & 0.032 & -0.219 & 0.732 & 0.01  &  -0.051 & 1 \\ [1ex] 
\hline 

\end{tabular}}

\label{cor}
\end{table}

The third calculation of the variance in $x^{T}$ is done by moving the target to its optimal location and orientation as determined by the minimization and then changing its position by a fixed translation in $x^{T}$. We restrict the minimization to translations parallel ($z^{T}$) and refit. This results in the columns of dots aligning in $x^{C}$, but misaligned in $z^{C}$. We find that a fixed translation in $x^{T}$ of 9.6 $\mathrm{\upmu m}$ increases the $\chi^{2}$ by 1 (corresponding to a one standard deviation $x^{T}$ translation). The three calculations of the variance produce comparable results. 

\begin{figure}[H]
{\centering
\includegraphics[width=8cm]{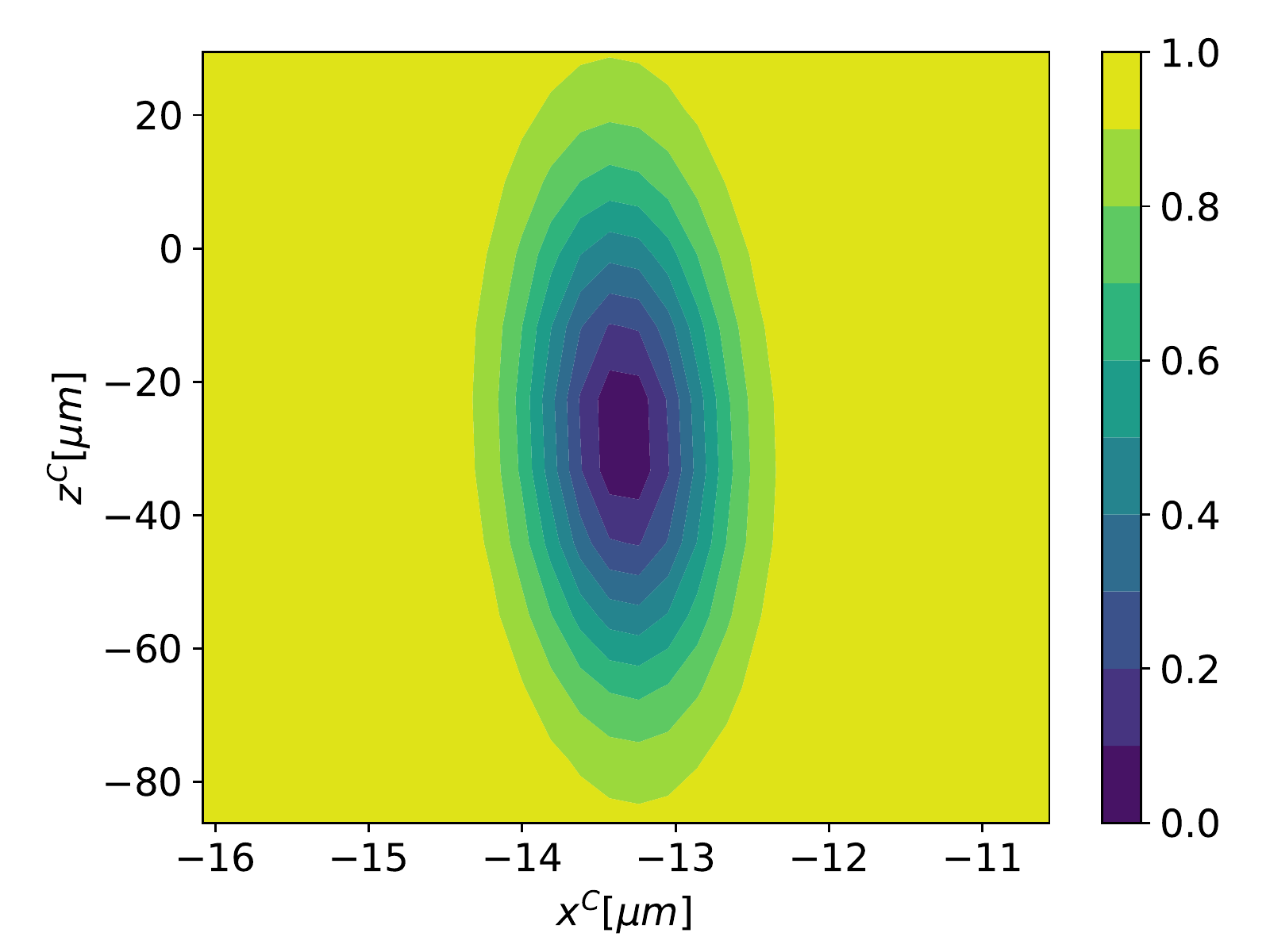}
\caption{Confidence region plot for parameters $x^{C}$ and $z^{C}$, produced by open source code\cite{LMFIT}.
\label{confidence}
}}\end{figure}

Additionally, an example confidence region plot for parameters $x^{C}$ and $z^{C}$ is shown in Figure \ref{confidence}, showing a minimal correlation between parameters $x^{C}$ and $z^{C}$, consistent with the small value of $Cor_{[x^{C}, z^{C}]}$ (Table \ref{cor}).

\begin{figure}[H]
{\centering
\includegraphics[width=8cm]{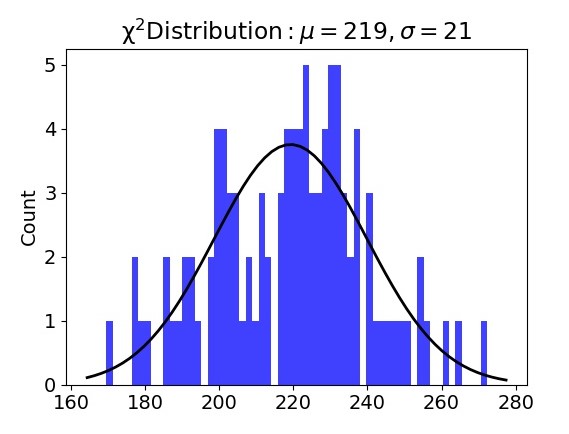}
\caption{The figure shows the $\chi^{2}_{2}$ distribution; the mean number of degrees of freedom is $\sim$ 220. The curve is a Gaussian fit to this distribution.
\label{chi_hist}
 }}\end{figure}

Finally, we show the distribution of the $\chi_{2}^{2}$ for $\sim $ 120 fits of images from a week of continuous data taking (Figure \ref{chi_hist}). The distribution does not include the fits using $\chi_{2}^{2}$ from the 20 images used to characterize the target. Here, the uncertainty is $\sigma = 0.12 $ pixels (described in Section \ref{image_analysis}), the mean of the distribution is $\sim 219$, whereas the mean number of degrees of freedom is $\sim 220$ indicating a good understanding of the precision of the measurement.

\subsection{Independent Axial Measurement} \label{indepz_section}

The correlation matrix and Figure \ref{confidence} demonstrate that $x^{C}$ translations are not significantly correlated with $z^{C}$ translations. Here we show an additional verification that the photographic technique's measurement of the axial coordinate, $z^{C}$, does not depend on the $x^{C}$ measurement by measuring $z^{C}$ completely independent of $x^{C}$. The procedure is to measure the magnification using pairs of vertically separated dots and using the magnification to infer the axial coordinate. 

Using the corrected magnification equation (Equation \ref{mag_eq}), the object distance is calculated for each column of dots using the two dots with the largest spacing. This $z^{C}$ coordinate for each column is then compared to the $z^{C}$ from that column in a reference image. The $z^{C}$ translation of the target is taken as the mean change in $z^{C}$ for the 17 pairs of dots.

\begin{figure}[ht]
{\centering
\includegraphics[width=8cm]{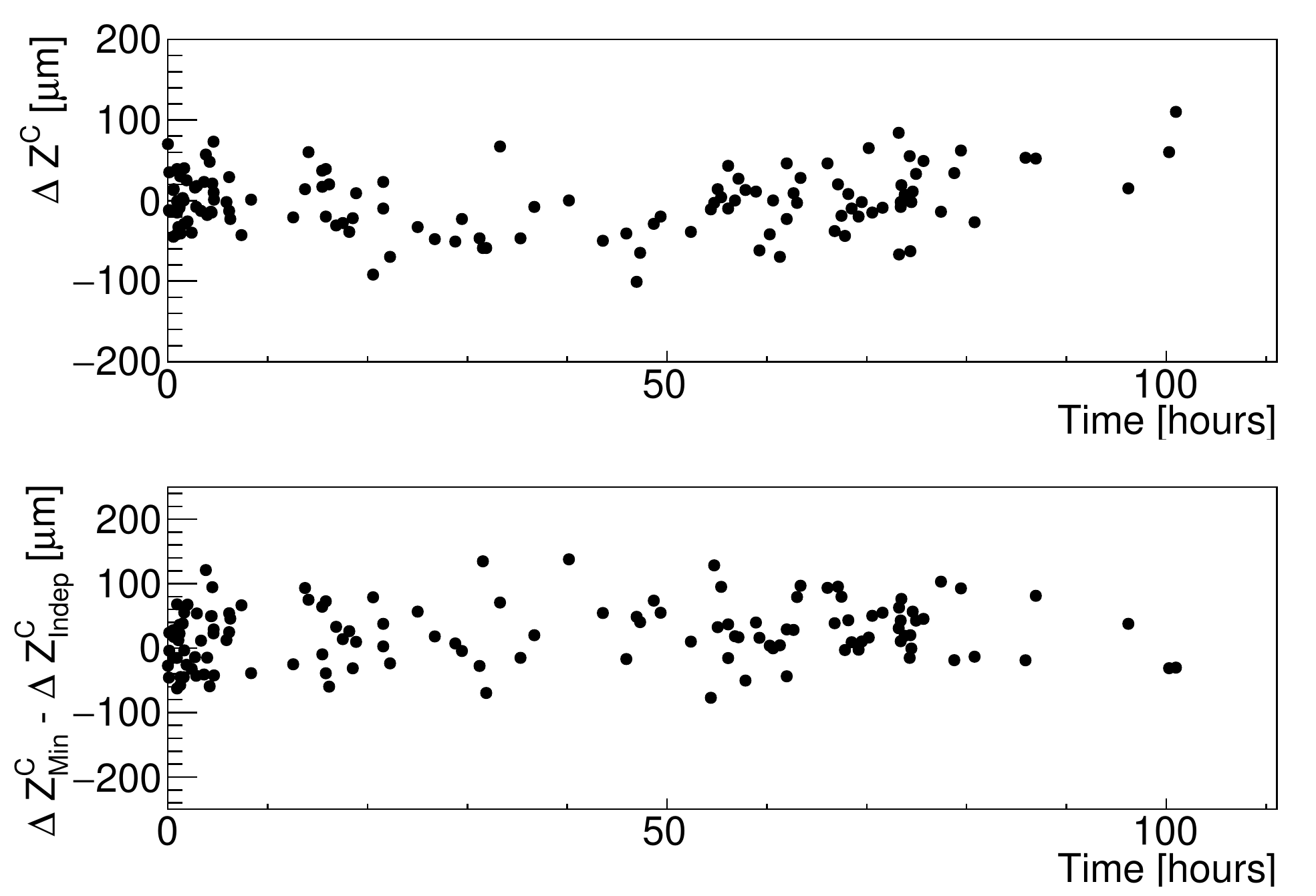}
\caption{Top: Translations in $z^{C}$ calculated from the full fit procedure as a function of time. Bottom: The difference in the $z^{C}$ translations calculated by the full fit procedure  and by the independent measurement.  The top graph shows a motion of the target while the bottom graph shows that the independent measurement of the motion tracks that of the full fit. The difference between the measurements is uniformly distributed ($\sigma \sim 50 $ $\mathrm{\upmu m}$) with an average difference of 21.7 $\upmu m$.
\label{indepz}
}}\end{figure}

Figure \ref{indepz} shows the difference in the $z^{C}$ translations as calculated by the independent measurement and the full fit. The $z^{C}$ translations calculated by the full fit are plotted for reference to show the time dependence of the fitted parameter. The figure demonstrates that the translation in $z^{C}$ measured by the independent measurement is consistent with the translation measured by the full fit.

\subsection{ Independent Measurement of Frame and Film} 
Finally, we provide an additional demonstration of the precision of the photographic technique by measuring the relative position of the frame and film planes normal to the target's surface ($x^{T}$). We calculate the {\it absolute} positions of the frame and film in the camera reference system using no information regarding their relative positions, and then we calculate the distance between the found positions. This distance is known from bench measurements of a few points on the frame to be $\sim$ 4.1 mm. This measurement has an uncertainty that we estimate as 100-200 $\upmu m$, largely due to the target not being fully characterized: the flatness of the target's frame has not been measured and there is evidence (Figure \ref{avg_res}) that the film contains a $\sim 100$ $\upmu m$ deformation. 

\begin{figure}[ht]
{\centering
\includegraphics[width=8cm]{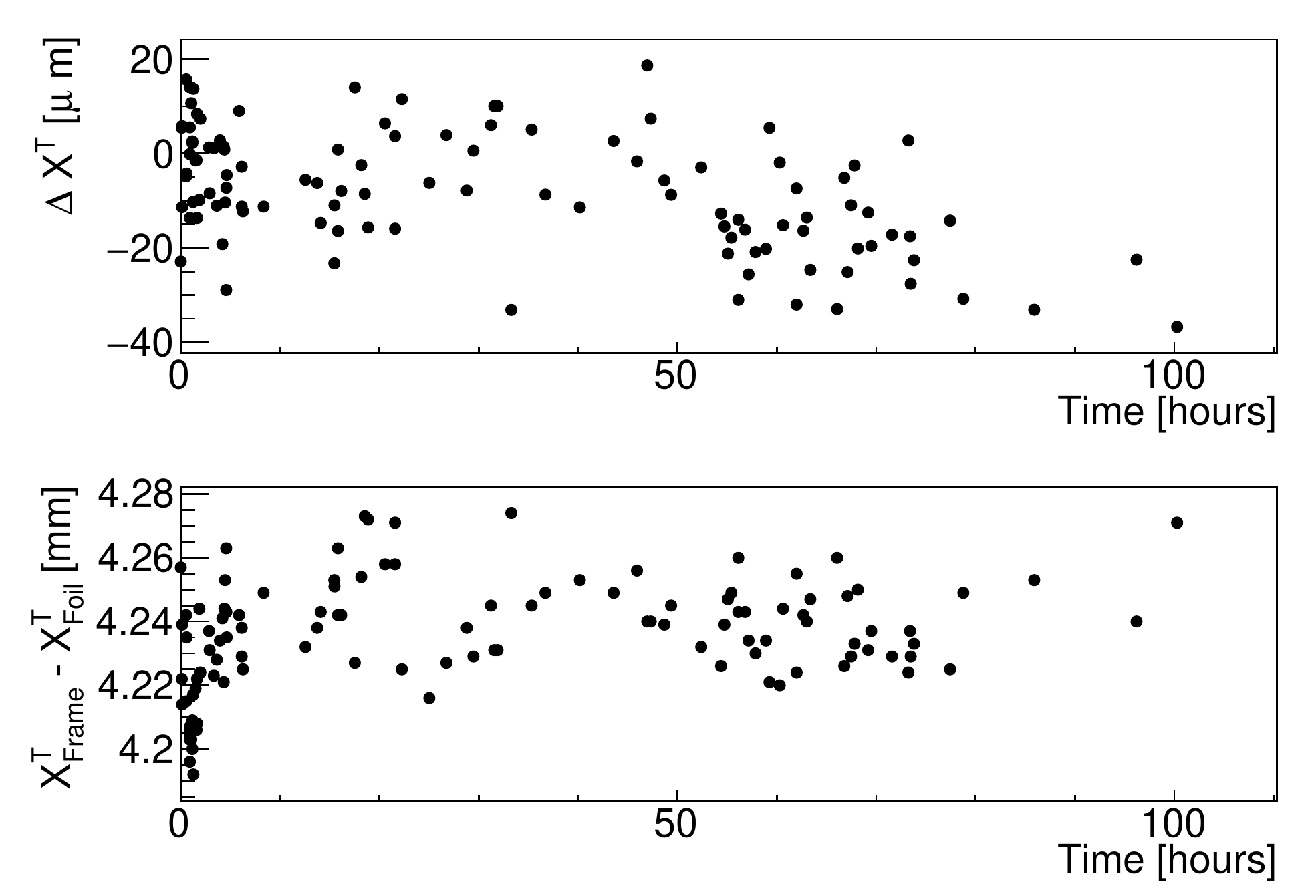}
\caption{Top: Translations in $x^{T}$ for the film as a function of time. Bottom: Difference between the frame and the film $x^{T}$ position showing good tracking with time.  
}\label{x_frame}}\end{figure}

The photographic technique calculates the average $x^{T}$ distance between the frame and film planes to be $4.24\pm 0.02$ mm. Since the film can deform to first order, the film's plane is defined by the film position at its edge, which is restricted to have a null deformation. The relative frame and film $x^{T}$ positions are plotted as a function of time in Figure \ref{x_frame}; the translations in $x^{T}$ for the film are plotted for reference to show the time dependence in the target position. The analysis measures minimal motion of the film with respect to the frame to the precision of this measurement, and the average distance is within the estimate of the uncertainty in the bench measurement. 
 
\section{Conclusions}
We have presented a noninvasive photographic technique to monitor, effectively continuously, the position, orientation, and shape of the MEG II muon stopping target, a thin film of scintillating plastic. Combined with infrequent bench measurements of the target geometry and optical surveys of the target position and the MEG II positron spectrometer, this technique will provide a measurement precision that exceeds the requirements for the experiment. We achieve a position uncertainty ($\sigma_{rms}$) of  $\sim$10 $\mathrm{\upmu m}$ normal to the target plane, $\sim$30 $\mathrm{\upmu m}$ horizontal and parallel to the surface of the film, and $\sim$1 $\mathrm{\upmu m}$ vertical and parallel to the surface of the film. These resolutions were achieved in an engineering run of the MEG II experiment, including operation in the magnetic field and radiation environment present during normal data taking. We have shown a number of consistency checks of the achieved resolution and examples of target motion and deformation found by the technique. Since the procedure only requires printed dots or other features on the surface to be monitored and an industrial camera, the procedure can be readily applied to any object that requires highly precise and continuous position monitoring.  

\section{Acknowledgements}
We thank the members of the MEG II collaboration for useful suggestions and help with some aspects of the implementation, particularly T. Iwamoto, P.-R. Kettle, A. Papa, F. Renga, S. Ritt, and C. Voena. We also thank UCI colleagues F. Badescu for help with the mechanics and electronics and G. Chanan for useful discussions on optics. We are grateful for the support and co-operation provided
by PSI as the host laboratory. This work was supported by U.S. Department of Energy grant DEFG02-91ER40679. 

\appendix
\section*{Appendices}
\addcontentsline{toc}{section}{Appendices}
\renewcommand{\thesubsection}{\Alph{subsection}}
\setcounter{table}{0}
\renewcommand{\thetable}{A\arabic{table}}
\setcounter{figure}{0}
\renewcommand{\thefigure}{A\arabic{figure}}

\subsection{Absolute Target Position Using Spectrometer Tracking} \label{app_tracking}
Here, we describe the technique used to check the position and orientation of the target with respect to the magnetic spectrometer independent of an optical survey. The technique is based on imaging small holes in the target using Michel positron trajectories measured in the magnetic spectrometer; it was first developed for MEG\cite{MEG}. This is an important part of the alignment procedure since it provides a check for potential errors in the optical survey of both the target and the tracking chambers in the magnetic spectrometer, in the magnetic field measurements, and in the particle tracking. The technique provides a measurement with very limited time dependence due to limited statistics, and can only measure the position of a limited number of holes and only near the target center where the muon stopping rate is high.  The photographic technique provides an effectively continuous monitoring of changes in the target position and shape over the full surface of the target to allow maximum use of the limited tracking statistics. 

The procedure is as follows. A set of small holes ($\sim 6$ mm diameter) will be made in the target film (6 were used in MEG). The positions of the holes will be imaged in 3 dimensions by detecting the deficit of positrons originating from the hole locations. The coordinate normal to the target surface is determined by measuring the apparent position of each hole in the target plane as a function of the angle of the positron as it intercepts the assumed target plane. We determine the position and direction of $e^{+}$ at the target by projecting the helical trajectory measured in the spectrometer back to the plane of the target\cite{MEG}. Target deformation and translation normal to the target surface are correlated with the linear first order dependence of each hole's apparent $y_{M}$ position on $tan{\phi}$, where $\phi$ is the $e^{+}$ angle at the target plane. The optimal target plane position is found when the apparent hole position is independent of angle.  

The MEG experiment used this correlation to measure each hole's position normal to the target surface with an uncertainty $\sigma$, which varied by year, ranging from 0.3-0.5 mm. This produced a systematic uncertainty in the signal acceptance that reduced the total sensitivity by 13\%\cite{MEG}. The uncertainty was largely the result of the lack of statistics available to measure the time dependence of the position. The photographic technique will allow time dependent corrections to the target position and allow the full tracking statistics to produce a high statistics check of the spectrometer-target relative alignment. This will allow full exploitation of the improved angular resolution of MEG II. 

\setcounter{section}{1}

\subsection{Threshold Parameter}\label{app_thresh}
\setcounter{table}{0}
\renewcommand{\thetable}{B\arabic{table}}
Here we confirm that the photographic technique is independent of the threshold parameter (described in Section \ref{image_analysis}).  First, we calculated the dispersion in individual dot positions in a single image while varying the threshold parameter from 50 to 100 (out of full scale value of 255).  The dispersion in the dot's position, $\sigma = 0.10$ pixels, is comparable  to the dispersion in sequential images taken with the same threshold parameter ($\sigma = 0.12$ pixels). 

Additionally, we analyzed 20 sequential images with varying threshold parameters in the range 50-90 and  calculated the dispersion in the values of the fit parameters for each image. The dispersion in the parameters for a given image with varying threshold parameter is comparable to the dispersion from sequential images with a constant threshold parameter, implying there are no systematic effects larger than the dispersion in the value of fitted parameters from sequential images. Both dispersion measurements are shown in Table \ref{thresh_table}. 

\begin{table}[ht]
\centering 
\caption{Top: The dispersion in the value of the fitted parameters for a given image with a varying threshold parameter (5 values ranging 50-90, out of full scale value of 255). Bottom: The dispersion in the value of the fitted parameters from sequential images with a constant threshold parameter is shown for reference.} \scalebox{0.75}{
\begin{tabular}{c c c c c c c c}
\hline\hline
$\phi$[mrad] & $\theta$[mrad] & $\psi$[mrad]  & $x^{C}$[$\mathrm{\upmu m}$]  & $y^{C}$[$\mathrm{\upmu m}$]  & $z^{C}$[$\mathrm{\upmu m}$]  & Bow[$\mathrm{\upmu m}$]  \\ [0.5ex]

\hline 
 0.02 & 0.01& 0.03 & 0.96 & 0.76 & 34.03 & 1.86\\ [1ex] 
\hline 
 0.02 & 0.01& 0.03 & 0.71 & 0.61 & 39.13 & 1.00 \\ [1ex] 

 \hline
\end{tabular}}
 
\label{thresh_table}

\end{table}

\subsection{Determination of Camera's Effective Focal Length}\label{app_optical}
\setcounter{figure}{0}
\renewcommand{\thefigure}{C\arabic{figure}}
\setcounter{table}{0}
\renewcommand{\thetable}{C\arabic{table}}
Here, we discuss the determination of the camera's effective focal length  ({\it efl}), which is related to the value of $I$ in Equation \ref{projection_eq} by the focal length approximation for an in-focus object. The manufacturer gives a nominal value of 50 mm for our complex lens. 

\begin{figure}[ht]
{\centering
\includegraphics[width=8cm]{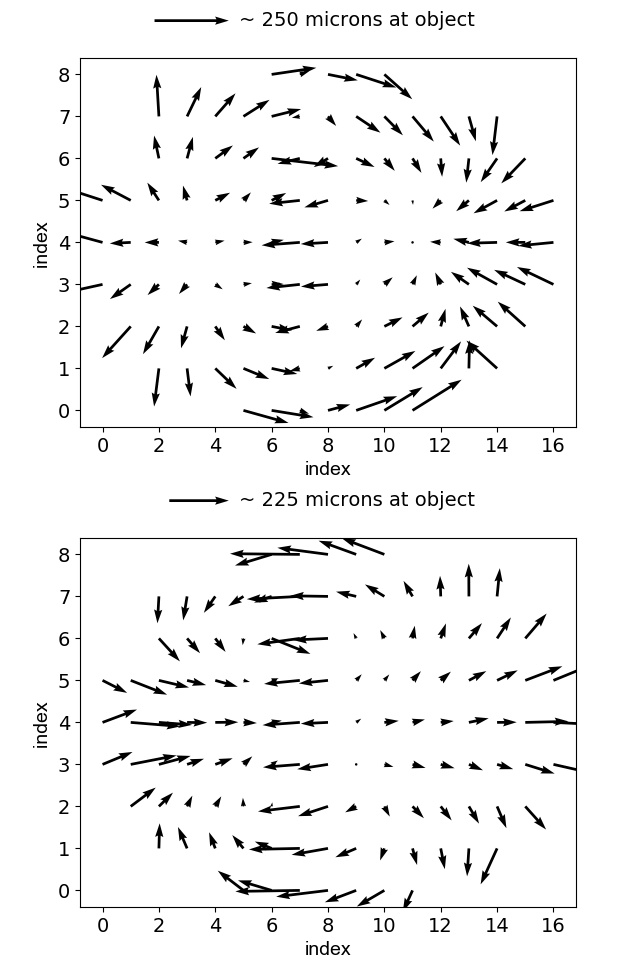}
\caption{The residuals are plotted as a function of their row and column indices.  The residuals on the top and bottom are from a minimization with a fixed $I$ of 48 mm and 54 mm respectively.  
\label{dipole}
}}\end{figure}

The {\it efl} is measured by exploiting the fact that an incorrect value for the {\it efl} creates characteristic systematic residuals for the dot positions in an object with significant depth of field. As an example, two residual plots with different fixed values of $I$ are shown in Figure \ref{dipole}. 

By including $I$ as an additional parameter in the fit, we reduce the magnitude of these residuals. We minimized the $\chi^{2}$ defined in Equation \ref{chi1_eq} (now with the additional parameter) on 20 sequential images to calculate the average optimal $I$. Dots with large systematic residuals (>10 $\sigma $), such as the $\sim 20$ dot deformation shown in Figure \ref{avg_res}, are excluded by doing a first fit, finding and removing dots with large residuals, and refitting. The optimal value of $I$ is found to be 51.61 $\pm$ 0.05 mm, corresponding to a best fit value of {\it efl} = 49.47 mm, close to the nominal 50 mm focal length given by the manufacturer. $I$ is fixed to this value for all analyses. 

Further, we verified that using a value of $I$ (and therefore the effective focal length) different than the best fit value does not affect the {\it change} in measured target position, orientation, and shape. We analyzed 20 sequential images with fixed $I$ values ranging from 49-53 mm. For each image in the set, the dispersion in the fit parameters as the value of $I$ is changed is significantly lower that of the dispersion from sequential images.

\end{document}